\documentclass[11pt]{article}
\usepackage{moriond,epsfig}

\begin{document}
\vspace*{4cm}
\title{RESULTS ON $B\to VV$ AND $PV$ DECAYS FROM BELLE}

\author{Jingzhi Zhang}

\address{KEK, High Energy Accelerator Research Organization, \\
1-1 Oho, Tsukuba, Ibaraki 305-0801, JAPAN}

\maketitle\abstracts{
  I report results on $B\to VV$ and $B\to PV$ decays.
  The results include the measurements of the decay
  amplitudes and the branching fractions in the decays $B\to \phi
  K^*$ and $B^+\to \rho^+\rho^0$, the measurements of the branching
  fraction and $CP$ asymmetry in $B^+\to \rho^+\pi^0$,
  and the first evidence of the decay $B^0\to\rho^0\pi^0$.}


\section{Introduction}

At the quark level, the decays $B \to \rho \pi$ occur via  $b\to u$
tree diagrams and can be used to measure the CKM angle
$\phi_2$~\cite{phi1-3}.
However, because of the presence of $b\to d$ penguin diagrams, the
extraction of $\phi_2$ from time-dependent $CP$-asymmetry measurements
requires an isospin analysis of the decay rates of all the $\rho \pi$
decay modes~\cite{iso_ana}. The decay channels $B^+ \to \rho^0 \pi^+$
~\cite{charge_conjugate} and $B^0 \to \rho^{\pm} \pi^{\mp}$ have
already been measured~\cite{current_rhopi}. The remaining decay modes,
$B^+\to \rho^+\pi^0$ and $B^0\to\rho^0\pi^0$, are reported here.
Direct $CP$ violation may occur in these decays because of
interference between the tree and penguin amplitudes. It would be
indicated by a non-zero partial-rate asymmetry: ${\cal A}_{CP}\equiv
\frac{{\Gamma{(\bar B \to \bar f)}-\Gamma{(B \to f)}}}{{\Gamma{(\bar B
      \to \bar f)}+\Gamma{(B \to f)}}}$, where $\Gamma{(B \to f)}$
denotes the partial width of $B$ decaying into a final state $f$ and
$\Gamma{(\bar B \to \bar f)}$ represents  that of the charge conjugate
decay.

In addition to rate asymmetries, $B\to VV$ decays provide
opportunities to search for direct $CP$ and/or $T$ violation through
angular correlations between the vector meson decay final
states~\cite{a.datta}. 
These decays produce final states where three helicity states are
possible.
The standard model (SM) predicts (1) $R_0\gg R_T=(R_\perp+R_\parallel)$, 
(2) $R_\perp \approx R_\parallel$~\cite{Grossman}, where $R_0$ $(R_T,
R_\perp, R_\parallel)$ is the longitudinal (transverse, perpendicular,
parallel) polarization fraction in the transversity
basis~\cite{ref_trans}.
In this report, we focus on the modes $B\to\phi K^*$~\cite{ref:phiK}
and $B^+\to\rho^+\rho^0$~\cite{ref:rhoprho0}.
The $B\to\phi K^*$ decays proceed via pure $b\to s$ penguin
diagrams, and are sensitive probes of new $CP$-violating phases from
physics beyond the SM~\cite{datta}.
The decay $B^+\to\rho^+\rho^0$ is a tree-dominated $b\to u$ process,
and can be used to extract $\phi_2$ by an isospin analysis analogous
to the $B\to \rho \pi$ decays.

The data samples, $140~{\rm fb}^{-1}$ used for the $\rho\pi$
modes and $78~{\rm fb}^{-1}$ for the $\phi K^{*}$ and
$\rho^+\rho^0$ modes, are collected with the Belle detector at the
KEKB asymmetric $e^+e^-$ collider~\cite{KEKB}. 
KEKB operates at the $\Upsilon(4S)$ resonance 
and has achieved a peak luminosity above $1.2 \times 10^{34}{\rm
  cm}^{-2}s^{-1}$.

\section{Event Selection}
We reconstruct $B$ meson candidates from their decay products
including the intermediate states $\phi \to K^+K^-$, 
$K^{* 0} \to K^{+} \pi^{-}$, 
$K^{* +} \to K^{+} \pi^{0}$,
$K^{* +} \to K^{0} \pi^{+}$,
$\rho^0 \to \pi^+\pi^-$, 
$\rho^+ \to \pi^+ \pi^0$ decays, 
and $\pi^0\to \gamma\gamma$ and $K^0\to K_S^0\to \pi^+\pi^-$.
$B$ candidates are identified using the beam-constrained mass $M_{\rm
  bc}\equiv\sqrt{E_{\rm beam}^2-p_B^2}$, and the energy difference
$\Delta E\equiv E_B-E_{\rm beam}$, where $E_{\rm beam}$ is the
center-of-mass system (CMS) beam energy, and $p_B$ and $E_B$ are the
CMS momentum and energy of the $B$ candidate, respectively.

The continuum process $e^+e^- \to q \bar{q}$ ($q=u, d, s, c$) is the
main source of background and must be strongly suppressed. 
One method of discriminating the signal from the background is based
on the event topology, which tends to be isotropic for $B\bar B$
events and jet-like for $q\bar q$ events.  
Another is $\theta_B$, the CMS polar angle of the $B$ flight
direction. $B$ mesons are produced with a $1-\cos^2\theta_B$
distribution while continuum background events tend to be uniform in
$\cos\theta_B$. 
We achieve continuum suppression by a likelihood ratio requirement
derived from a Fisher discriminant based on modified Fox-Wolfram
moments~\cite{fox} and $\theta_B$.


\section { $VV$ Modes: $B\to \phi K^*$, $B^+\to \rho^+ \rho^0$}

The $B\to \phi K^*$ signal yields are extracted by $2D$ extended unbinned
maximum-likelihood fits to the $\Delta E$-$M_{\rm bc}$ distributions.
The non-resonant $B \to KKK^{(*)}$ background is estimated from the
$\phi$ sideband region and is subtracted from the raw signal
yield. The branching fractions are  
$$ {\cal B}(B\to \phi
K^{*0})=(10.0^{+1.6}_{-1.5}\;^{+0.7}_{-0.8})\times
10^{-6},~~ 
{\cal B}(B \to \phi
K^{*+})=(6.7^{+2.1}_{-1.9}\;^{+0.7}_{-1.0})\times10^{-6},$$
where the first (second) error is statistical (systematic) throughout
this paper.

\begin{figure}[htbp]
\begin{center}
\includegraphics[width=3.5cm, clip]{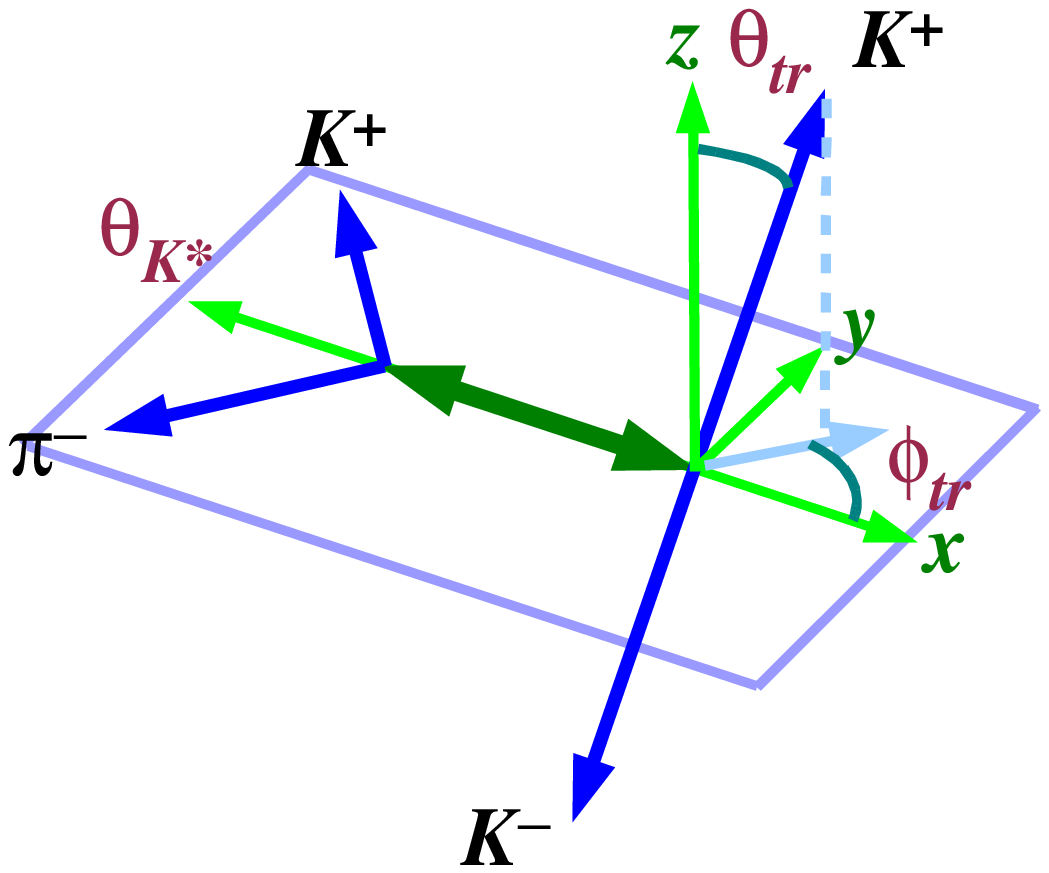}
\includegraphics[width=8.1cm, clip]{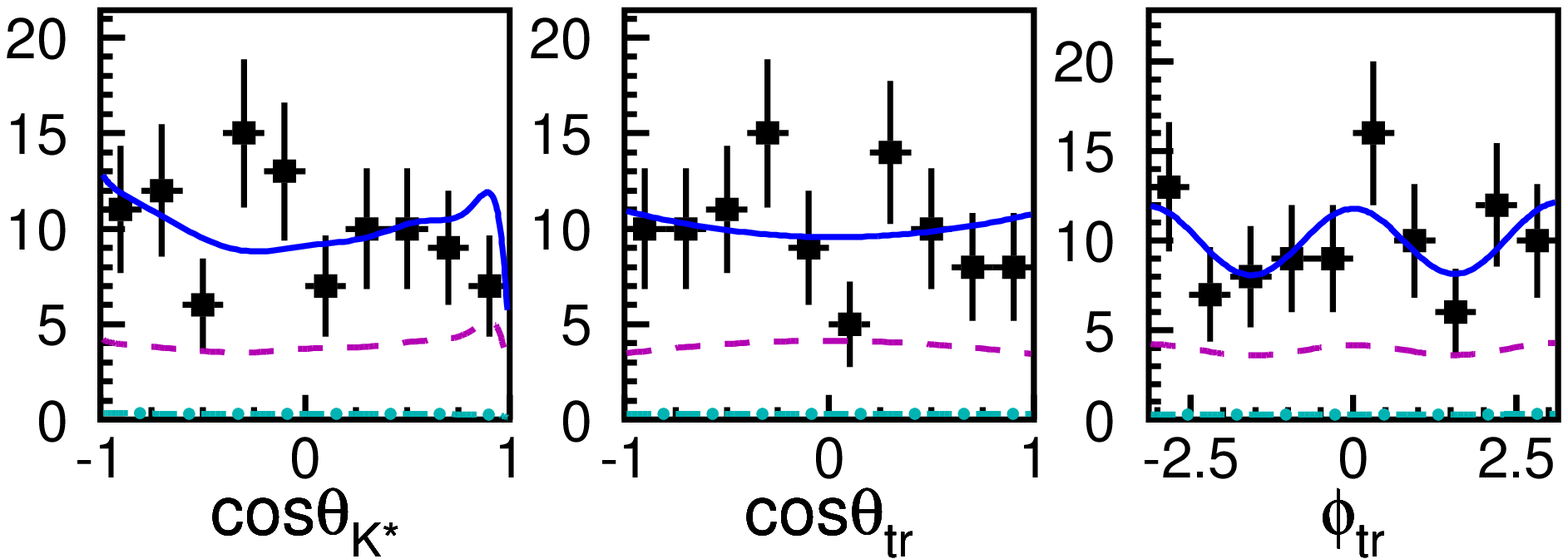}
\caption{(a) the definition of decay angles in $B \to
  \phi K^{*0}$ decay; (b $\sim$ d) the projections of the angles with results
  of the fit superimposed, the dashed (dot-dashed) line denotes the
  continuum ($B\to KKK^*$) background.}
\label{fig_phiK}
\end{center}
\end{figure}

The decay angles of $B \to \phi K^{*0}(K^+\pi^-)$ are defined in the
transversity basis, as shown in Fig.~\ref{fig_phiK} (a). 
The distribution of the three angles $\theta_{K^*}$, $\theta_{\rm
  tr}$, and $\phi_{\rm tr}$ is  \cite{polarization_hepex}
 \begin{eqnarray}
\label{equ:angularpdf}
\nonumber
&&{d^3 \Gamma (\phi_{\rm tr}, \cos\theta_{\rm tr}, \cos\theta_{K^*}) \over d\phi_{\rm tr} d\cos{\theta_{\rm tr}} d\
cos{\theta_{K^*}}} 
= {9\over 32\pi} [ 
|A_\perp|^2 2 \cos^2{\theta_{\rm tr}} \sin^2{\theta_{K^*}} 
+|A_\parallel|^2 2 \sin^2{\theta_{\rm tr}} \sin^2{\phi_{\rm tr}}
\sin^2{\theta_{K^*}}\\
\nonumber
&&+{|A_0|}^2 4 \sin^2\theta_{\rm tr} \cos^2\phi_{\rm tr} \cos^2\theta_{K^*} 
+\sqrt{2} {\rm Re}(A^*_\parallel{A_0}) \sin^2\theta_{\rm tr}\sin 2 \phi_{\rm tr} \sin 2\theta_{K^*} \\
&&-\eta\sqrt{2} {\rm Im}({A_0}^*  A_\perp) \sin 2\theta_{\rm tr} \cos\phi_{\rm tr} \sin 2\theta_{K^*} 
-2\eta{\rm Im}(A_\parallel^* A_\perp) \sin 2 \theta_{\rm tr} \sin \phi_{\rm tr} \sin^2 \theta_{K^*} ]~,
\end{eqnarray}
where $A_0$, $A_\parallel$, and $A_\perp$ are the complex amplitudes
of the three helicity states in the transversity basis with the
normalization condition $|A_0|^2 + |A_\parallel|^2 + |A_\perp|^2 = 1$,
and $\eta\equiv +1$ ($-1$) for $B^0$ ($\overline{B}{}^0$). 


The complex amplitudes are determined from an unbinned maximum
likelihood fit to the $B^0 \to \phi K^{*0}$ candidates. The
combined likelihood is given by
$\mathcal{L} =
\prod_i^N \epsilon[f_{\phi K^{*0}}\cdot\Gamma
+f_{q\overline{q}}\cdot P_{q\overline{q}}
+ f_{KKK^{*0}}\cdot P_{KKK^{*0}} ],$
where $\Gamma$ is the angular distribution function (ADF) given in
Eq.~\ref{equ:angularpdf}, and $R_{q\overline{q}}$ and $R_{KKK^{*0}}$
are the ADFs for continuum and $B \to KKK^{*0}$ background. 
$R_{q\overline{q}}$ is determined from sideband data and
$R_{KKK^{*0}}$ is assumed to be flat. 
The detection efficiency function $\epsilon$ is determined by Monte
Carlo (MC).
The fractions of $\phi K^{*0}$ ($f_{\phi K^{*0}}$), $q\overline{q}$
($f_{q\overline{q}}$) and $KKK^{*0}$ ($f_{KKK^{*0}}$) are
parameterized as a function of $\Delta E$ and $M_{\rm bc}$.
Four parameters ($|A_0|^2$, $|A_\perp|^2$, $\arg(A_\parallel)$,
$\arg(A_\perp)$) are left free;
$\arg(A_0)$ is set to zero and $|A_\parallel|^2$ is calculated from
the normalization constraint.
Projections of the three angles with fit results are shown in 
Fig.~\ref{fig_phiK} (b) $\sim$ (d).\\

\vspace*{0.2cm}
\begin{tabular}{c c c c}
\hspace*{0.5cm}(a)\hspace*{4.3cm}&(b)\hspace*{1.7cm}&(c)\hspace*{1.7cm}& (d)
\end{tabular}
\vspace*{-1.2cm}

The amplitudes obtained from the fit are
\begin{eqnarray}
\nonumber
 |A_0|^2 = 0.43 \pm 0.09\pm 0.04;~~&&
 |A_\perp|^2 = 0.41 \pm 0.10\pm 0.04;\\
\nonumber
 \arg(A_\parallel) = -2.57 \pm 0.39\pm 0.09;~~&&
 \arg(A_\perp) = 0.48 \pm 0.32\pm 0.06.
\end{eqnarray}

\begin{figure}[htbp]
\begin{center}
\includegraphics[width=3.6cm, clip]{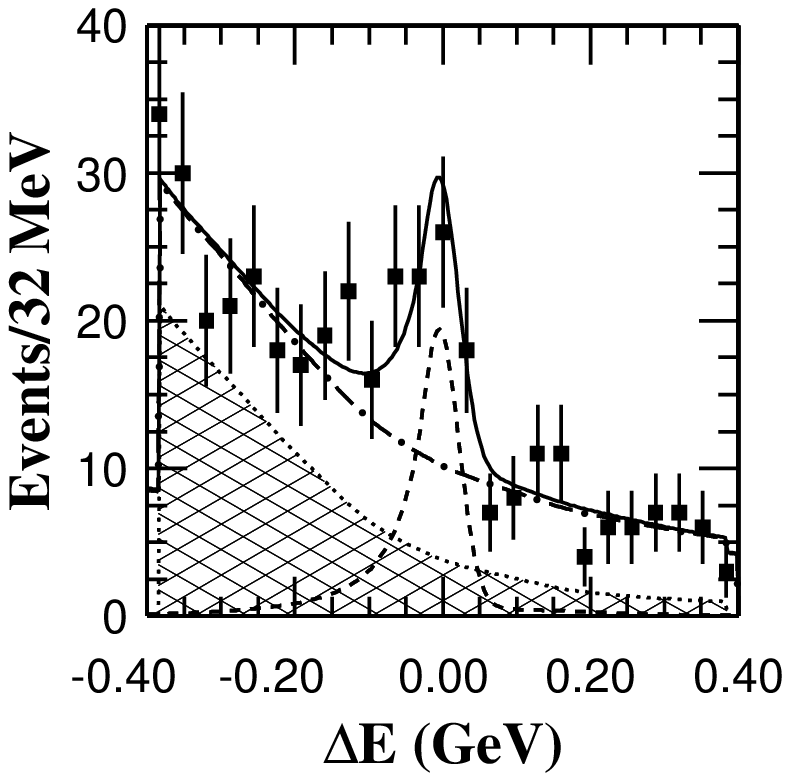}
\includegraphics[width=3.6cm, clip]{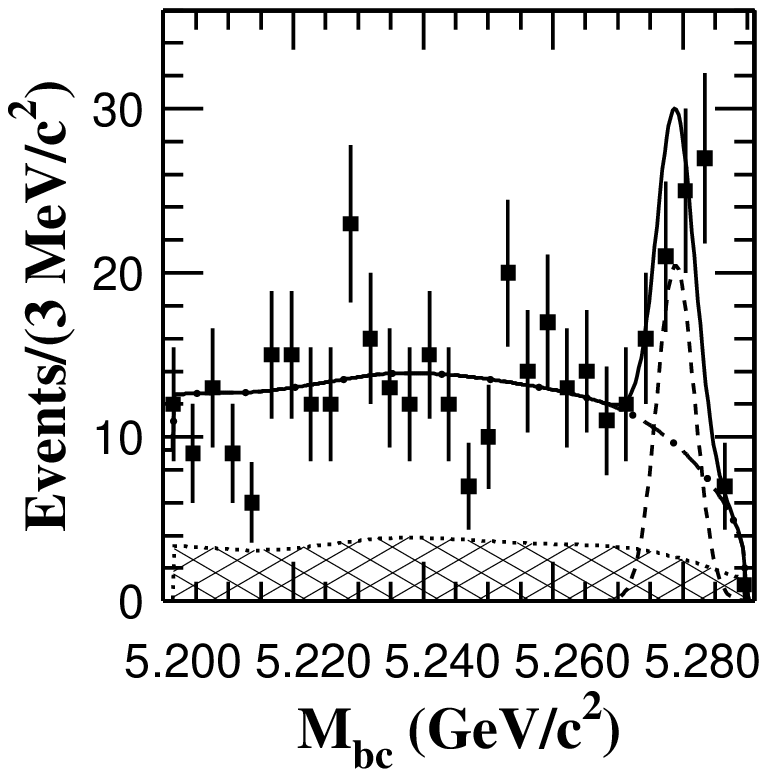}
\includegraphics[width=3.6cm, clip]{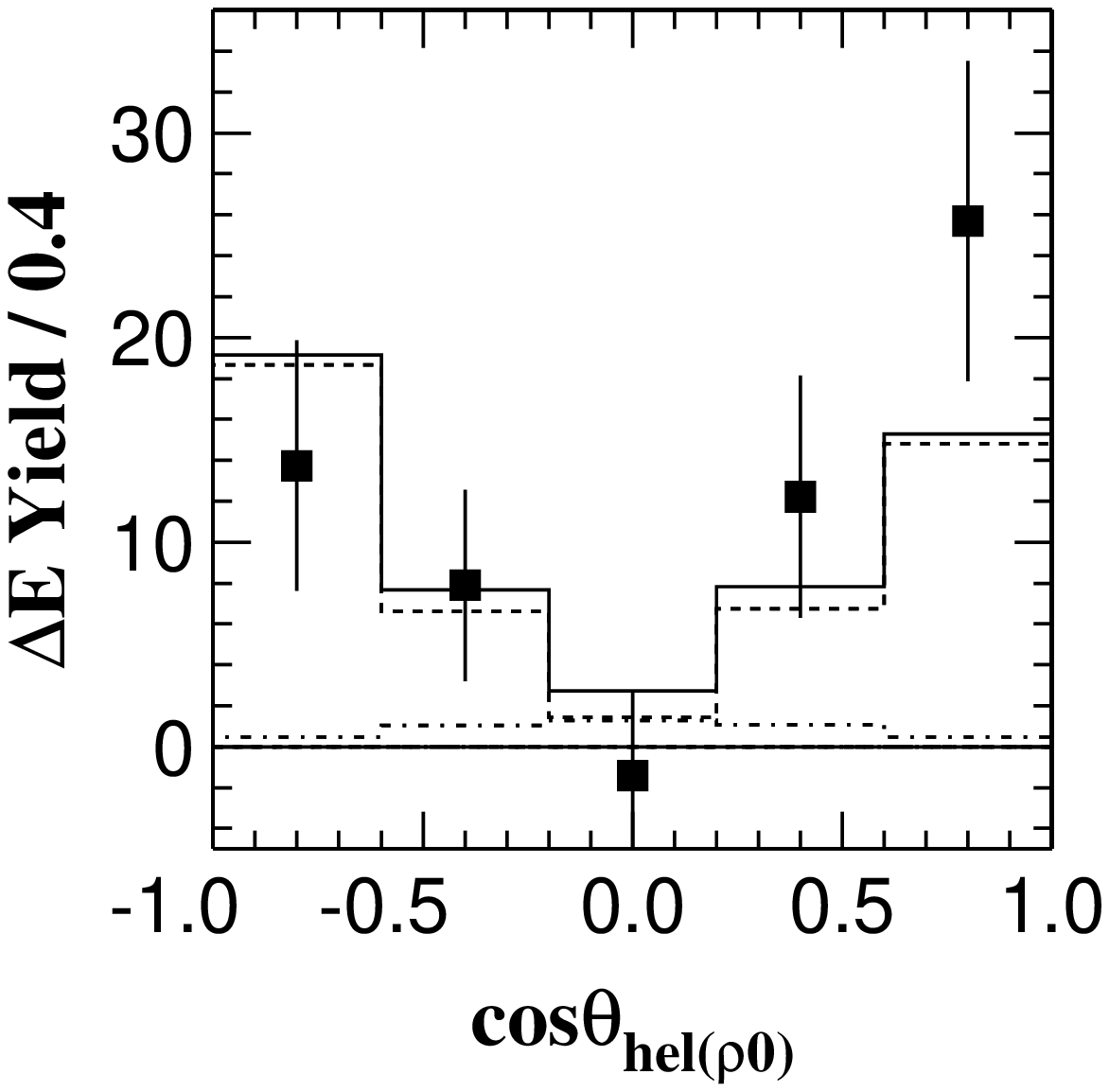}
\includegraphics[width=3.6cm, clip]{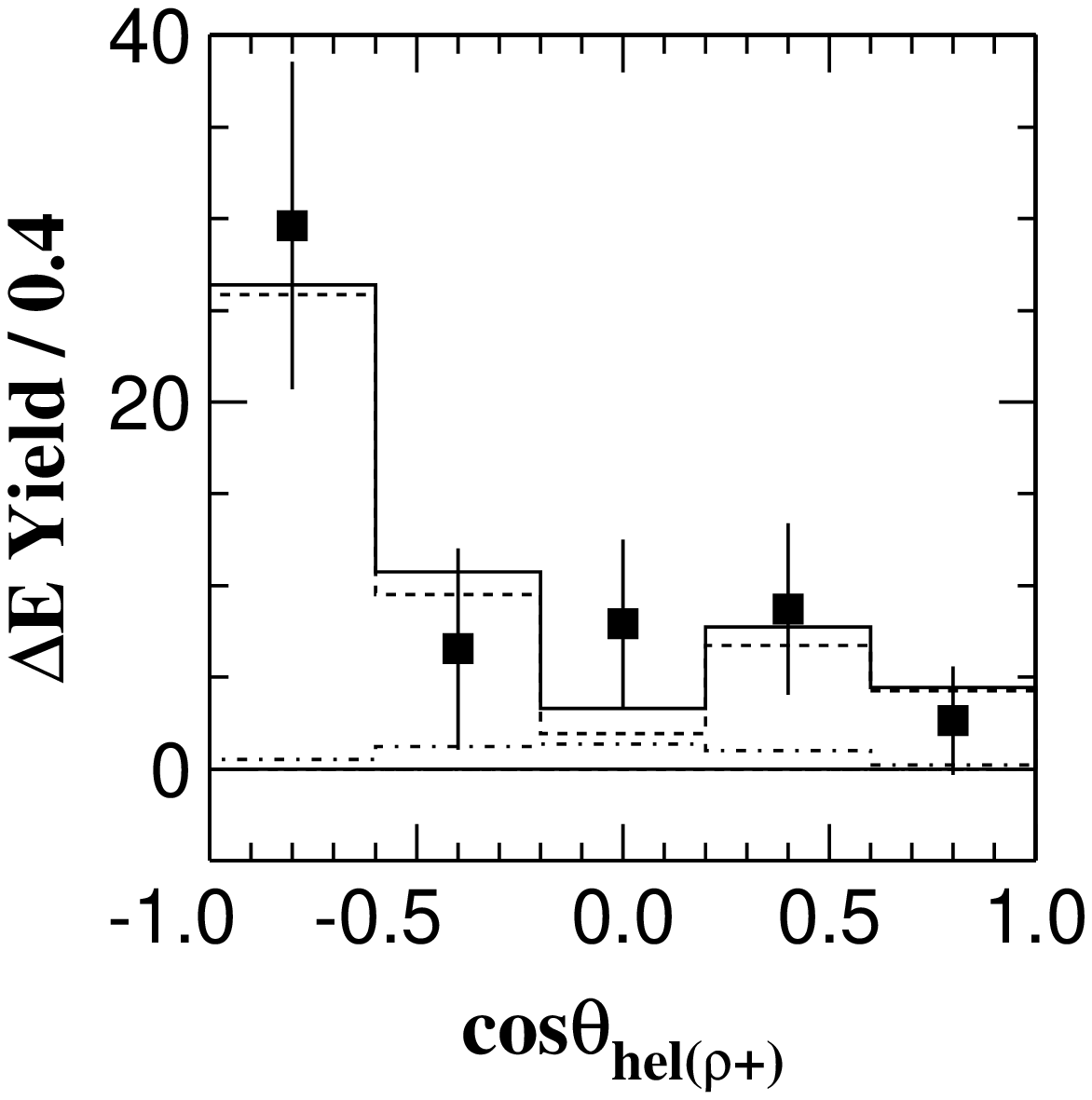}
\caption{(a) the $\Delta E$ and (b) $M_{\rm bc}$ fits to the $B^+\to
  \rho^+\rho^0$ candidate events.
  The sum of the $b\to c $ and continuum components is shown as a
  dashed line;  shaded histogram represents the $b\to c$
  background. (c)(d): the data points show the background-subtracted
  cosine helicity-angle distributions  $\rho^0$ (c) and $\rho^+$
  (d). The dashed (dot-dashed) histogram is the $H_{00}$ ($H_{11}$)
  component of the fit; the solid histogram is their sum. The absence
  of events near $\cos\theta_{hel_{\rho^+}}=1$ is due to a $\pi^0$
  momentum requirement $p_{\pi^0}>0.5{\,{\rm GeV}/c}$.}
\label{fig_rhorho}
\end{center}
\end{figure}

Figures~\ref{fig_rhorho} (a) and (b) show the $\Delta E$ and $M_{\rm bc}$
projections for $B^+\to \rho^+\rho^0$. The curve shows the results of
a binned maximum-likelihood fit with three components: signal,
continuum, and $b\to c$ background. The $\Delta E$ fit
gives a signal yield of $59 \pm 13$ entries. The statistical
significance of the signal, defined as $\sqrt{-2\ln({\cal
    L}_0/{\cal L}_{\rm max})}$, where ${\cal L}_{\rm max}$ is the
likelihood value at the best-fit signal  yield and ${\cal L}_{0}$ is
the value with the signal yield fixed to zero, is 5.3$\sigma$.

We use the $\rho\to\pi\pi$ helicity-angle ($\theta_{\rm {hel}}$)
distributions to determine the relative strengths of the longitudinally
and transversely polarization. Here $\theta_{\rm {hel}}$ is the angle
between an axis anti-parallel to the $B$ flight direction and the
$\pi^+$ flight direction in the $\rho$ rest frame. 
The signal yields for each helicity-angle bin are plotted versus
$\cos\theta_{\rm hel}$ for the $\rho^0$ in Fig.~\ref{fig_rhorho} (c) and
the $\rho^+$ in Fig.~\ref{fig_rhorho} (d). 
We perform a simultaneous $\chi^2$ fit to the two
$\rho$ helicity-angle distributions using
MC-determined  expectations for the longitudinal and transverse helicity
states. The fit results are shown as histograms in
Fig.~\ref{fig_rhorho} (c) and(d). 
Since the detection efficiency is strongly dependent on
polarization, we calculate the branching fraction  based on the measured
longitudinal polarization fraction $R_0$ (note that
$R_0=\frac{|A_0|^2}{|A_0|^2 + |A_\parallel|^2 + |A_\perp|^2}$),

\begin{eqnarray}
\nonumber
R_0(B^+\to \rho^+\rho^0)&=& 0.95\pm 0.11\pm 0.02,\\
\nonumber
{\cal B}(B^+\to \rho^+\rho^0)&=&(31.7\pm 7.1^{+3.8}_{-6.7}\;)\times
10^{-6}.
\end{eqnarray}

\vspace*{0.2cm}
\begin{tabular}{c c c c}
\hspace*{0.6cm}(a)\hspace*{2.8cm}&(b)\hspace*{2.8cm}&(c)\hspace*{2.8cm}& (d)
\end{tabular}

\vspace*{-1.2cm}

We see that in the tree-dominated $B\to\rho^+\rho^0$, the SM
prediction~\cite{Grossman} $R_0 \gg R_T$ is confirmed.
The second prediction, $R_\perp \approx R_\parallel$, cannot be tested
at the current level of statistics.
In contrast, in the pure $b\to s$ penguin $B\to\phi K^*$ we find
$R_0\approx R_T$; also find $R_T\gg R_\parallel$
($R_0+R_\perp+R_\parallel=1$). Both of these results for $B\to\phi
K^*$ are in disagreement with SM predictions.

\section{$PV$ Modes: $B^+\to \rho^+\pi^0$, $B^0\to \rho^0\pi^0$}

From the ${\rm pseudoscalar} \to {\rm vector}+{\rm pseudoscalar}$ decay
$B\to \rho\pi$, we expect the $\rho$ helicity
angle ($\theta_{\rm hel}$) to have a $\cos^2\theta_{\rm hel}$
distribution. 
We apply the following requirements:  $|\cos \theta_{\rm
  hel}|>0.3$ for $B^+\to \rho^+\pi^0$ and $\left|
\cos\theta_{\rm hel} \right| > 0.5$ for $B^0\to \rho^0\pi^0$.
Additional discrimination is provided by the $b$-flavor tagging
parameter $r$, which is a measure of the probability that the $b$
flavor of the accompanying $B$ meson is correctly assigned
by the Belle flavor-tagging algorithm~\cite{Belle-sin2phi1}. 
Events with a high value of $r$ are well-tagged and are less likely to
originate from continuum events. We extract signal yields by using
extended unbinned maximum-likelihood fits to the $M_{\rm bc}$-$\Delta
E$ distributions.

\begin{figure}[htbp]
\begin{center}
\includegraphics[width=7cm, clip]{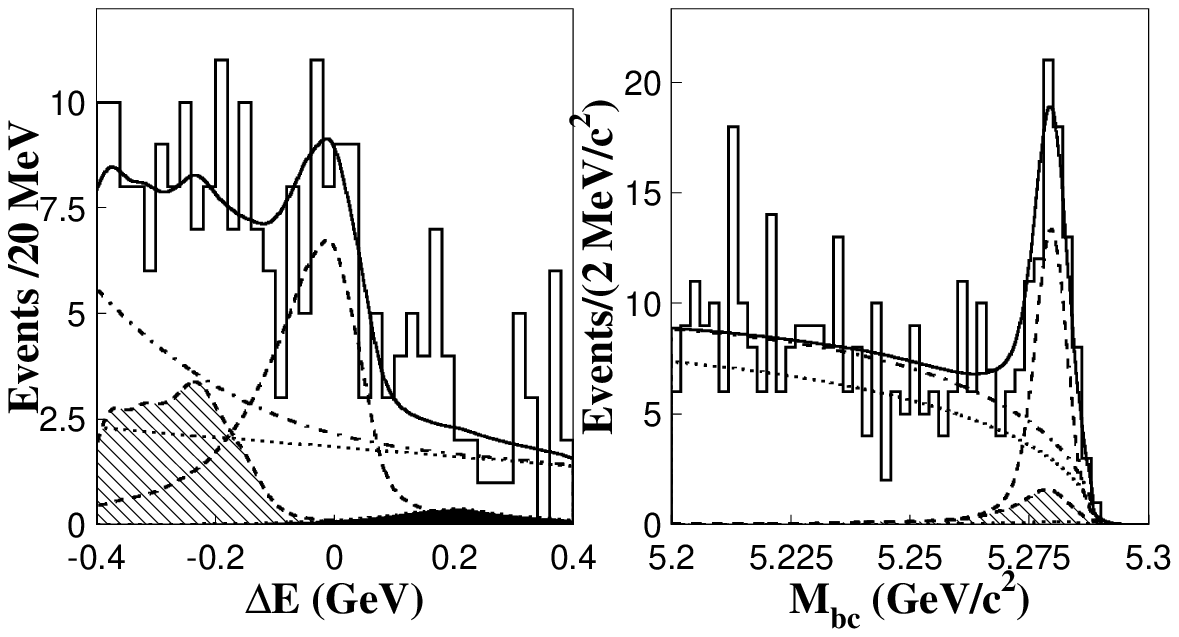}
\includegraphics[width=7cm, clip]{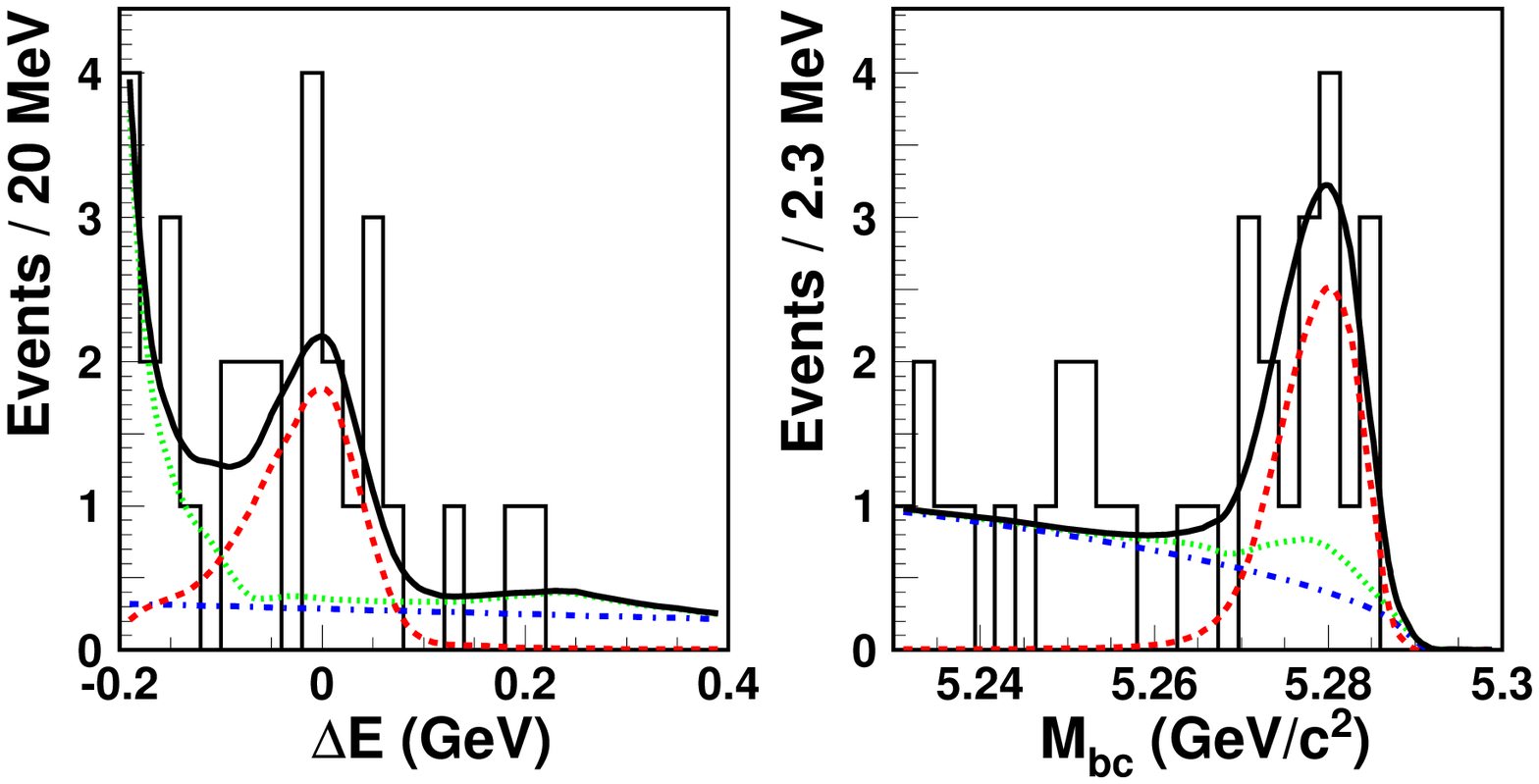}
\caption{ 
  (a) the  $\Delta E$ projection in the $M_{\rm bc}$ 
  signal region, (b) the $M_{\rm bc}$  projection in the $\Delta E$
  signal region for the decay $B^+\to \rho^+\pi^0$. The
  solid curve shows the fit results. The signal (continuum, the
  sum of continuum and  $b\to c$) component is shown as dashed
  (dotted, dot-dashed) line. The hatched (dark) histogram represents
  the $B\to \rho^+\rho^-$ ($B\to \pi^0\pi^0$) background;
  (c)(d) for $B^0\to \rho^0\pi^0$, the solid curve is a projection
  of the maximum likelihood fit result. The dashed (dot-dashed,
  dotted) curve  represents the signal (continuum, the composite of
  continuum and  B-related background) component of the fit.}
\label{fig_rhopi}
\end{center}
\end{figure}

Figures~\ref{fig_rhopi} (a) and (b) show the $\Delta E$ and $M_{\rm bc}$
projections for $B^+\to \rho^+\pi^0$.
The solid curve shows the fit results with the components: 
signal, continuum, the $b\to c$ decays, $B^0 \to \rho^+\rho^-$ and
$B^0\to\pi^0\pi^0$.
In the fit, all normalizations are allowed to float, except for the
$\pi^0\pi^0$ component, which is fixed at a MC-determined value based
on recent Belle~\cite{shlee} and BaBar~\cite{BaBar_pi0pi0} measurements.
The fit gives a signal yield of $87 \pm 15$, with a statistical
significance of 8.1$\sigma$.\\

\begin{tabular}{c c c c}
\hspace*{1cm}(a)\hspace*{2.4cm}&(b)\hspace*{2.8cm}&(c)\hspace*{2.6cm}& (d)
\end{tabular}

\vspace*{-1.1cm}

Figures~\ref{fig_rhopi} (c) and (d) show the fit results for $B^0\to
\rho^0\pi^0$.
The fit contains components for the signal, continuum, $b
\to c$ background  and the decays $B^+ \to \rho^+ \rho^0$, $B^+ \to
\rho^+ \pi^0$ and $B^+ \to \pi^+ \pi^0$. 
The normalizations of the $B^+ \to \rho^+\pi^0$ and $B^+ \to \pi^+\pi^0$ 
components are fixed according to previous
measurements~\cite{current_rhopi}, while the normalizations
of all other components are allowed to float.
The signal yield is found to be $15 \pm 5$ with 3.6$\sigma$ significance.


We use a simultaneous fit to extract the partial rate asymmetry (${\cal
  A}_{CP}$) by introducing asymmetry parameters into the
  $B^\mp\to\rho^\mp\pi^0$  fit.
The measured ${\cal A}_{CP}$ together with the branching fractions
are summarized in Table.~\ref{table_rhopi}.
\vspace*{-0.3cm}
\begin{table}[htbp]
  \caption{Signal yields ($N_{\rm sig}$), significance ($S$), efficiencies
  ($\epsilon$), branching fractions and ${\cal A}_{CP}$
  \label{table_rhopi}}
\begin{center}
\begin{tabular}{c|c|c|c|c|c}\hline
 Modes  & $N_{\rm sig}$ & $S$ & $\epsilon$ &  Branch Fraction($\times 10^{-6}$) &  ${\cal A}_{CP}$\\ \hline
$B^+\to\rho^+\pi^0$ & $87\pm15$ & 8.1 & 4.4\%  & $13.2\pm 2.3^{+1.4}_{-1.9}\;$ & $0.06\pm0.19\pm0.04$\\ \hline
$B^0 \to \rho^0\pi^0$ & $15\pm 5$ & 3.6 & 1.91\%  & $5.1 \pm 1.6\pm 0.8$ & -\\ \hline
\end{tabular}
\end{center}
\end{table}

\vspace*{-0.2cm}
\section*{Summary}
In summary, we measured the branching fractions of the decays $B\to\phi
K^*$, $B^+\to \rho^+\rho^0$. We observed the decay $B^+\to \rho^+\pi^0$,
and the first evidence for $B^0 \to \rho^0\pi^0$.
An angular analysis is performed on the $VV$ modes.
It indicates that, in the tree-dominated decay $B^+\to \rho^+\rho^0$, the
longitudinal polarization is saturated ($R_0\approx 1$), which is
consistent with SM predictions. 
However, in the pure $b\to s$ penguin decay $B\to\phi K^*$, $R_0$ and
$R_T$ are comparable, while $R_\perp$ is significantly larger than 
$R_\parallel$;
these results are in disagreement with SM predictions.
It is thus important to obtain polarization measurements in
other modes, especially the pure penguin $b\to s \bar d d$ decay,
$B^+\to K^{*0}\rho^+$.


\section*{References}
\begin{thebibliography}{99}

\bibitem{phi1-3}
H.~R.~Quinn and A.~I.~Sanda, Eur. Phys. Jour. C{\bf 15}, 626 (2000).

\bibitem{iso_ana}
H.~J.~Lipkin, Y.~Nir, H.~R.~Quinn, A.~Snyder, Phys. Rev. D{\bf 44},
1454 (1991);\\
A.~E.~Snyder, H.~R.~Quinn, Phys. Rev. D{\bf 48}, 2139 (1993).

\bibitem{charge_conjugate}
The inclusion of charge conjugate modes is implied unless stated otherwise.
 
\bibitem{current_rhopi}
A.~Gordon {\it et al.} (Belle Collaboration), Phys. Lett. B{\bf 542}, 183 (2002);\\
B.~Aubert {\it et al.} (BaBar Collaboration), Phys. Rev. Lett. {\bf 91}, 201802 (2003);\\
B.~Aubert {\it et al.} (BaBar Collaboration), hep-ex/0311049, submitted to Phys. Rev. Lett.

\bibitem{a.datta}
A.~Datta, D.~London, hep-ph/0303159 (2003).

\bibitem{Grossman}
Y.~Grossman, Int.~J.~Mod.~Phys. A{\bf 19}, 907 (2004).

\bibitem{ref_trans}
I.~Dunietz, H.~Quinn, A.~Snyder, W.~Toki, and H.~J.~Lipkin,
Phys. Rev. D{\bf 43}, 2193 (1991).

\bibitem{ref:phiK}
K.~F.~Chen {\it et al.} (Belle Collaboration), Phys. Rev. Lett. {\bf
  91}, 201801 (2003).

\bibitem{ref:rhoprho0}
J.~Zhang {\it et al.} (Belle Collaboration), Phys. Rev. Lett. {\bf
  91}, 221801 (2003).

\bibitem{datta}
A.~Datta, Phys. Rev. D{\bf 66}, 071702, (2002).

\bibitem{KEKB}
S.~Kurokawa and E.~Kikutani, Nucl. Instr. Meth. A{\bf 499}, 1 (2003).

\bibitem{fox}
G.~C.~Fox, S.~Wolfram, Phys. Rev. Lett. {\bf 41}, 1581 (1978);\\
K.~Abe {\it et al.} (Belle Collaboration), Phys. Rev. Lett. {\bf 87},
101801 (2001).

\bibitem{polarization_hepex}
K.~Abe, M.~Satpathy and H.~Yamamoto, hep-ex/0103002 (2001).

\bibitem{Belle-sin2phi1}
K.~Abe {\it et al.} (Belle Collaboration), Phys. Rev. D{\bf 66},
071102 (2002).

\bibitem{shlee}
S.~H.~Lee {\it et al.} (Belle Collaboration), Phys. Rev. Lett. {\bf
  91}, 261801 (2003).

\bibitem{BaBar_pi0pi0}
B.~Aubert {\it et al.} (BaBar Collaboration), Phys. Rev. Lett. {\bf
  91}, 241801 (2003).

\end{thebibliography}
\end{document}